\documentclass[a4paper,notitlepage]{report}   	%
\renewcommand{\thechapter}{{\textbf{\arabic{chapter}}}}
\renewcommand{\thesection}{\thechapter:$\,$\arabic{section}}

\renewcommand{\appendix}{
	\section*{Appendices}
	\setcounter{section}{0}
	\renewcommand{\thesection}{\thechapter:$\,$\Alph{section}}
}
\newcommand{\appendixend}{
	\setcounter{section}{0}
	\renewcommand{\thesection}{\thechapter:$\,$\arabic{section}}
}
\usepackage{amsmath}
\usepackage{amssymb}
\usepackage{xr}
\usepackage{refcount}

\usepackage{graphicx}
\usepackage{mathrsfs}
\usepackage{bm}
\usepackage[hyperref]{xcolor} %
\definecolor{darkgreen}{rgb}{0,0.4,0} %
\definecolor{darkred}{rgb}{0.55,0,0} %
\definecolor{navy}{rgb}{0,0,0.55} %
\usepackage[hypertexnames=false,breaklinks=true,colorlinks=true,citecolor=darkgreen,linkcolor=black,urlcolor=navy]{hyperref} %
\usepackage{amsthm}
\theoremstyle{definition} %
\newtheorem*{lemma}{Lemma}
\newtheorem*{lproof}{Proof}
\newtheorem*{definition}{Definition}

\newcommand{\G}{\mc{G}}

\newcommand{\prm}[1]{\protect{$#1$}}

\newcommand{\sol}[1]{}
\newcommand{\h}{\mathrm{H}}%
\newcommand{\bmh}{{\bm{\mathrm{H}}}}%

\newcommand{\ta}{{\tilde{a}}}

\newcommand{\ulambda}{\lambda} %
\newcommand{\mrm}[1]{\mathrm{#1}}
\newcommand{\mbf}[1]{\mathbf{#1}}
\newcommand{\mbb}[1]{\mathbb{#1}}
\newcommand{\mc}[1]{\mathcal{#1}}

\newcommand{\mscr}[1]{\mathscr{#1}}
\newcommand{\bgrid}{\left(\begin{array}{rrr}}
\newcommand{\egrid}{\end{array}\right)}
\newcommand{\bgridt}{\left(\begin{array}{rr}}
\newcommand{\egridt}{\end{array}\right)}
\newcommand{\bgridtt}{\left[\begin{array}{rr}}
\newcommand{\egridtt}{\end{array}\right]}
\newcommand{\os}{\!\!\not\!} %
\newcommand{\eref}[1]{(\ref{#1})}
\newcommand{\Eref}[1]{Eq.~(\ref{#1})}

\newcommand{\Erefr}[2]{Eqs.~(\ref{#1}--\ref{#2})}

\newcommand{\erefr}[2]{(\ref{#1}--\ref{#2})}

\newcommand{\erefs}[2]{(\ref{#1}) and~(\ref{#2})}

\newcommand{\cref}[1]{Chapter~\ref{#1}}
\newcommand{\Cref}[1]{Chapter~\ref{#1}}

\newcommand{\Peref}[2]{(\ref{#2})}
\newcommand{\PEref}[2]{Eq.~(\ref{#2})}

\newcommand{\sref}[1]{Sec.~\ref{#1}}
\newcommand{\Psref}[2]{\sref{#2}}

\newcommand{\srefs}[2]{Secs.~\ref{#1} and~\ref{#2}}

\newcommand{\srefr}[2]{Secs.~\ref{#1}--\ref{#2}}
\newcommand{\aref}[1]{Appendix~\ref{#1}}

\newcommand{\bsm}{\bar\sigma^\mu}
\newcommand{\bsmm}{\bar\sigma_\mu}

\newcommand{\nn}{\nonumber}
\newcommand{\rmd}{\mathrm{d}}
\newcommand{\rmi}{\mathrm{i}}%
\newcommand{\la}{\langle}
\newcommand{\ra}{\rangle}

\newcommand{\Dslash}{{\,\os\! D}}
\newcommand{\bmDslash}{\bm{\Dslash}}

\newcommand{\Tr}{\mrm{Tr}\,}

\newcommand{\rcite}[1]{Ref.~\cite{#1}}

\newcommand{\p}[1]{\phantom{#1}}

\newcommand{\fg}{\mrm{fg}}
\newcommand{\bgfield}[1]{[#1]_\mrm{bg}}
\newcommand{\fgfield}[1]{[#1]_\fg}
\newcommand{\OO}[1]{\mrm{O}\left(#1\right)}

\newcommand{\SU}[1]{\mrm{SU}(#1)}
\newcommand{\su}[1]{\mrm{su}(#1)}

\newcommand{\U}[1]{\mrm{U}(#1)}

\newcommand{\rmU}{\mrm{U}}

\newcommand{\pEref}[1]{\protect{Eq.~(\ref{#1})}}
\newcommand{\perefr}[2]{\protect{(\ref{#1}--\ref{#2})}}

\newcommand{\pcref}[1]{\protect{Chapter~\ref{#1}}}

\newcommand{\GL}[2]{\mrm{GL}(#1,\mbb{#2})}
\newcommand{\GLp}[2]{\mrm{GL}^+(#1,\mbb{#2})}
\newcommand{\gl}[2]{\mrm{gl}(#1,\mbb{#2})}
\newcommand{\SL}[2]{\mrm{SL}(#1,\mbb{#2})}

\newcommand{\GLTC}{\GL{2}{C}}%

\renewcommand{\bar}[1]{\overline{#1}}

\newcommand{\preon}{\psi}

\newcommand{\thps}{\theta_\psi}

\newcommand{\ILO}[1]{\mrm{O}(#1)}

\newcommand{\bmmf}{\tilde{f}}

\newcommand{\N}{{\mathfrak{n}}}%
\newcommand{\RM}{{{\mbb{R}^{1,3}}}}

\newcommand{\Cw}[1]{{{\mbb{C}^{\wedge #1}}}}

\newcommand{\vp}{\varphi}
\newcommand{\tvp}{{\tilde{\varphi}}}
\newcommand{\bmvp}{{\bm{\vp}}}

\newcommand{\Cwt}{{\Cw{2}}}

\newcommand{\Cwtn}{{\Cw{2\N}}}
\newcommand{\Cws}{{\Cw{6}}}

\newcommand{\Ptagref}[2]{{\tag{\ref{#2}}}}%

\newcommand{\quietGversion}[1]{}
\newcommand{\onlyinsummary}[1]{}

\newcommand{\completed}[1]{}

\newcommand{\chap}[1]{#1}
\newcommand{\notchap}[1]{}
\usepackage{geometry}
\usepackage[numbers,square,sort&compress]{natbib}

\begin{document}

\title{A Classical Analogue to the Standard Model\\and General Relativity}
\author{Chapter 2\\~\\Robert N. C. Pfeifer}

\date{12 June 2023}

\maketitle
\thispagestyle{empty}
\newpage

\tableofcontents

\setcounter{chapter}{1}
\chapter{Colour interactions and fermions from scalar fields on \protect{$\Cw{6}$}\label{ch:colour}}

\begin{abstract}
This chapter continues the study of quasiparticles on complex manifolds with anticommuting co-ordinates, and shows that on increasing the dimensionality of the complex manifold from $\mathbb{C}^{\wedge 2}$ to $\mathbb{C}^{\wedge 6}$, the dimension-$L^{-1/2}$ spinor excitations in the associated effective field theory on $\mathbb{R}^{1,3}$ acquire an $\mathrm{SU}(3)$ colour charge and assemble into composite spinors of dimension $L^{-3/2}$. This model provides a novel environment in which to study colour confinement, while also demonstrating the ability of the first member of the $\mathbb{C}^{\wedge 6\mathfrak{n}}$ series to support fermions having spin and dimensions consistent with the Standard Model.
\end{abstract}

\section{Introduction}

\notchap{Quasiparticles are ubiquitous in both condensed matter and particle physics. In condensed matter they} 
\chap{Quasiparticles in condensed matter physics} 
tend to take the form of collective excitations in many-body systems, such as phonons in an acoustically condictive medium \cite{srivastava1990} or solitons in a fluid \cite{drazin1989}. More speculatively, this category could also include exotic topological phenomena such as skyrmions \cite{skyrme1961,skyrme1962,tokura2020}, hopfions \cite{gobel2020}, hedgehogs \cite{tokura2020,gobel2020}, and anyons \cite{kitaev2006,bonderson2007}, with growing evidence for their experimental observation in %
many body systems \cite{nakamura2020,bartolomei2020}. In particle physics all the mesons are composite and behave as particles only in the effective field theory of the (comparatively) low-energy regime \cite{gell-mann1964,zweig1964a}. It has been speculated that the quarks and leptons may themselves also be composite, with constituents termed \emph{preons} \cite{dsouza1992}, though no experimental evidence for such structure has yet been observed.

\cref{ch:simplest} %
introduced a quasiparticle model on a manifold with anticommuting co-ordinates denoted $\Cwt$. Although individual fields on this manifold are not normalisable, they may collectively support normalisable excitations which behave as quasiparticles under a mapping to $\RM$. The species exhibited in this model are limited, comprising one vector boson, one complex scalar boson and its antiparticle, and a pair of spin-$\frac{1}{2}$ fermions with dimension $L^{-{1}/{2}}$. To move beyond this, the present chapter extends the model on $\Cwt$ to obtain the series $\Cwtn$. Specialising to a particular member of that series, $\Cw{6}$, the fermions of dimension $L^{-{1}/{2}}$ are then seen to carry a colour charge mediated by an $\SU{3}$ gauge field. These fermions behave as a simple system of preons, supporting the construction of several species of spin-$\frac{1}{2}$ fermions of dimension $L^{-{3}/{2}}$. Like a Cooper pair assembled from dressed electrons, or holes, these emergent fermions are %
assembled from preons which are themselves many-body excitations of a condensed matter system. %

\section{Mapping from \protect{$\Cw{2\lowercase{\mathfrak{n}}}$} to \protect{$\RM$}\label{sec:II:mapping}}

\subsection{Conventions}

This chapter follows the same conventions as \cref{ch:simplest}, including frequent use of units in which $h=c=1$. 

Given a Riemann manifold $M$, the notation $TM$ denotes the tangent bundle to $M$ and $T^*M$ denotes the cotangent bundle to $M$. If $E$ is a Lie group then the notation $E\otimes T^*M$ denotes an $E$-valued cotangent bundle %
on $M$. Writing $e_a$ for an orthonormal basis of $E$, a 1-form on $E\otimes T^*M$ may be written $e_a\phi_\mu^a \,\rmd x^\mu$, and is associated with an $E$-valued covector field $e_a\phi_\mu^a $ on $M$. 

For clarity, the ``dual'' of a space always refers to the relationship between a space of $k$-vectors (for specific $k$) and the corresponding space of $p$-forms. If it is necessary to discuss both the dual of a space and the conjugate of that same space, these are referred to as the dual space and the conjugate space respectively.

When equations or lemmas from \cref{ch:simplest} are referenced, these take the form (\textbf{1}.1). %

The symbols $\Upsilon^\mu$, $\bar\Upsilon^\mu$, and $\tilde\Upsilon^\mu$ satisfy
\begin{align}
\bar\Upsilon^\mu\Upsilon^\nu&=\eta^{\mu\nu}\Ptagref{I}{eq:defUpsilon1}\\
\tilde\Upsilon^\mu\tilde\Upsilon^\nu&=\eta^{\mu\nu}.\Ptagref{I}{eq:bmDtilde}
\end{align}

To avoid confusion with supersymmetry $R$-indices and other general roman counting indices, when referring to a model in the series $\Cwtn$ the parameter of the series is written as $\N$, not $n$.

When enumerating a basis of matrices, indices which are not identifiable with holomorphic (undotted) or antiholomorphic (dotted) co-ordinates on $\Cwtn$ are written with a tilde, $\ta$.

\subsection{Symmetries of \protect{$\Cwtn$}---revisited\label{sec:C2nsym}}

As observed in %
\srefs{sec:symCwn}{sec:repCwt}, the vector space $\Cwtn$ is a $2\N$-dimensional vector space whose basis elements takes values in a $2\N$-dimensional tier-1 subset of the Grassmann algebra. That is, where $\zeta_i$ is an anticommuting generator of the Grassmann algebra,
\begin{equation}
\Lambda_\infty\leftarrow \bm\{1,\{\zeta_i|i\in\mbb{Z}^+\};\cdot\,;+;\mbb{C}\bm\},
\end{equation}
the basis vectors of $\Cwtn$ correspond to $\{\zeta_i|i\in\mbb{Z}_{2\N}\}$. These vectors then inherit an anticommuting product operation from $\Lambda_\infty$ when constructing $k$-vectors with $k>1$, such that (for example)
\begin{equation}
\begin{split}
\mbf{v}&=v^i\zeta_i\qquad \mbf{w}=w^i\zeta_i\\
\mbf{A}&:=\mbf{v}\mbf{w}=-\mbf{w}\mbf{v}\\&\p{:}=\frac{1}{2}(v^iw^j-v^jw^i)\zeta_i\zeta_j.
\end{split}
\end{equation}
The space of co-ordinates on $\Cwtn$ is symmetric under the action of $\GL{2\N}{C}$, and also under the %
$2\N$-dimensional translations on $\Cwtn$, denoted $T^{2\N}_\Cwtn$. Co-ordinates on $\Cwtn$ are anticommuting $2\N$-component complex vectors, written $\theta_{\bm{\alpha}}$ with index in bold. The symmetry groups $\GL{2\N}{C}$ and $T^{2\N}_\Cwtn$ act on these vectors by matrix multiplication and vector addition respectively.

To introduce a metric on $\Cwtn$, it is first convenient to introduce a new form for co-ordinate vector $\theta_{\bm{\alpha}}$. To do so, observe that by Lemma~\eref{Lem:4} %
the $\GL{2}{C}$ symmetry of $\Cwt$ decomposes as
\begin{equation}
\GL{2}{C}\cong\SL{2}{C}\oplus \GL{1}{C},\label{eq:II:GL2Cdecomp}
\end{equation}
where $\GL{1}{C}$ is the multiplicative group of complex numbers, also denoted $\mbb{C}^\times$. %
Similarly, using %
Lemmas~\eref{Lem:4}, \eref{Lem:1} and~\eref{Lem:9} of \aref{apdx:lemmas}
the $\GL{2\N}{C}$ symmetry of $\Cwtn|_{\N>1}$ may be rewritten %
\begin{align}
\!\!\!\!
\GL{2\N}{C}
&\cong\GL{\N}{R}\otimes\GL{2}{C}\label{eq:GL2NCdecomp1}\\
&\cong \left[\SU{\N}\oplus\GL{1}{R}\right]\otimes\left[\SL{2}{C}\oplus\mbb{C}^\times\right].
\label{eq:GL2NCdecomp2new}
\end{align}
From \Eref{eq:GL2NCdecomp1}
it follows that $\Cwtn$ admits a system of co-ordinates comprising $\N$ two-component complex vectors where %
each such vector parameterises a $\Cwt$ subspace. %
It is then convenient to replace the $2\N$-component vector $\theta_{\bm{\alpha}}$ with the two-index object $\theta_{a\alpha}$ where $a$ is %
acted on by the $\GL{\N}{R}$ symmetry and ranges from 1 to $\N$, and $\alpha$ is acted on by $\GL{2}{C}$ and ranges from 1 to 2. Much as $\Cwt$ can be embedded in $\mbb{R}^{4|4}$, being a subspace of the supersymmetric extension of a point in the $\RM$ submanifold, manifold $\Cwtn$ can be embedded over a point in $\mbb{R}^{4|4\N}$ such that index $a$ is the index of an $R$-symmetry. %
Invariance under $\GL{\N}{R}$ implies that $\delta^{ab}$ is a natural choice of metric on the %
$R$-indices, and for vectors written as $\theta_{a\alpha}$ the metric on $\Cwtn$ takes the form 
\begin{equation}
g^{a\alpha\,b\beta}=\varepsilon^{\alpha\beta}\delta^{ab}.
\end{equation}

To identify an $\RM$ submanifold on which to study the emergent quasiparticle theory, arbitrarily select the $\Cwt$ subspace defined by
\begin{equation}
\theta^{a\alpha}=\theta^{1\alpha}%
\end{equation}
in some connection-free but otherwise arbitrarily-chosen co-ordinate system having form $\theta_{a\alpha}$ and covering $\Cwtn$.
Let this specific subspace be denoted $\Cw{2\bullet}$, and recognise that there are an infinite number of equivalent subspaces
\begin{equation}
\{a\Cw{2\bullet}~|~a\in\GL{\N}{R}\}.
\end{equation}
The choice of subspace $\Cw{2\bullet}$ from within this set is arbitrary. As in \sref{sec:anticommmanif}, this subspace is a double cover of $\RM$, and a mapping $\G$ may be introduced which is 1:1 between a subspace of $\Cw{2\bullet}$ and all of $\RM$. As in \sref{sec:defpullback}, %
let $M\subset\Cwtn$ denote $\G^{-1}(\RM)$. Co-ordinates $x^\mu$ may then equally well denote points on $\RM$ or on $M$ as required.

\subsection{Field content on \protect{$\RM$}\label{sec:fieldcontentRM}}

It then remains to see how the presence of an additional $\GL{\N}{R}$ symmetry acting on the $R$-indices affects the emergent quasiparticle model present on $\RM$ in the low-energy limit. %
First, 
recall that the effective fields of the quasiparticle model arise from the gradients of products of fields on $\Cwtn$. With $2\N$ anticommuting co-ordinates the Taylor series of the component fields contain more terms than on $\Cwt$ but nevertheless still terminate after a finite number of terms and the construction remains the same. As mapping $\G$ is projective over the 
$R$-indices, the resulting product field $\vp(x)$ remains single-valued. However, given a point $P\in\RM$ and its image $\G^{-1}(P)\in\Cwtn$, the family of gradient fields at $\G^{-1}(P)$ %
becomes richer with increasing $\N$. As before, each linearly independent derivative operator on $\Cwtn$ may be associated with a linearly independent field on $\RM$.

First there are the chiral fields which are associated with chiral derivative operators,
\begin{align}
\psi^{a\alpha}(x)\qquad&\partial^{a\alpha}\label{eq:Cwtnfields1}\\
\bar\psi_{\dot a\dot\alpha}(x)\qquad&\bar\partial_{\dot a\dot\alpha}.
\end{align}
There are now $\N$ holomorphic and $\N$ antiholomorphic spinor fields of dimension $L^{-1/2}$, each carrying one $R$-index [acted on by $\GL{\N}{R}$] and one $S$-index [acted on by $\GLTC$].

Next, there are the vector fields and their associated derivative operators. As indicated in \sref{sec:cauchy}, there are now $\N^2$ vector derivatives of $\vp(x)$ at $\G^{-1}(P)$ and these define $\N^2$ vector fields
\begin{align}
\left.\vp^{\dot aa}_\mu(x)\,\right|_\RM:=\left.\partial^{\dot aa}_\mu\,\vp[\G^{-1}(x)]\,\right|_\Cwtn\qquad\partial^{\dot aa}_\mu:=\bar\partial^{\dot a}\bsmm\partial^a. \label{eq:vpaa}
\end{align}
These may equivalently be written as a single $\gl{\N}{R}$-valued vector field $e_{\dot aa}\vp_\mu^{\dot aa}(x)$.

Finally, there is again a complex scalar sector. At first glance it appears that this sector also has a basis of $\N^2$ such fields, $\{\partial^{m\alpha}\partial^n_\alpha\vp(x)\}$. However, anticommutation and index raising/lowering yield
\begin{equation}
\partial^{m\alpha}\partial^n_\alpha\vp(x)=\partial^{n\alpha}\partial^m_\alpha\vp(x),
\end{equation}
reducing the number of independent fields to $(\N^2+\N)/2$. Further, this restriction is not invariant under the $\GL{\N}{R}$ symmetry of \Eref{eq:Cwtnfields1}. To symmetrise with respect to $\GL{\N}{R}$, recognise that there exist global transformations in $\GL{\N}{R}$ whose associated Lie algebra homomorphisms map any two such scalar fields into positions $(m,n)$ and $(n,m)$, and thus all scalar fields of the form $\partial^{m\alpha}\partial^n_\alpha\vp(x)$ are equal. It therefore suffices to consider only the explicitly $\GL{\N}{R}$-invariant scalar field $\h(x):=\partial^{n\alpha}\partial_{n\alpha}\vp(x)$,
\begin{equation}
\begin{split}
\h(x)\qquad&\partial_\rmU:=\partial^{a\alpha}\partial_{a\alpha}
\label{eq:Cwtnfields5}
\end{split}
\end{equation}
with conjugate
\begin{equation}
\begin{split}
\h^*(x)\qquad&\bar{\partial}_\rmU:=\bar\partial_{\dot a\dot\alpha}\bar{\partial}^{\dot a\dot\alpha}.
\end{split}
\end{equation}

To study these quantities as quasiparticle fields on $\RM$ it is necessary to confirm that each has a unique associated quantity which behaves as its space--time derivative, and then to ensure that anticommutation once again yields an appropriately-constructed Lagrangian for each field, permitting extrapolation of these fields from their values on an initial Cauchy surface. For $\N=1$ the derivative operator was constructed as per \Eref{eq:mudiv}, with natural generalisation
\begin{equation}
\bm{\partial}_\mu:=e_{\dot aa}\bar\partial_{\dot\alpha}^{\dot a}\bar\sigma^{\dot\alpha\alpha}
_{\mu}\partial_{\alpha}^a.\label{eq:genmudiv}
\end{equation}
However, for the vector fields $\vp^{\dot aa}_\mu(x)$, linear independence implies that a derivative $\partial^{\dot bb}_\mu\vp^{\dot aa}_\nu(x)$ may always be rewritten
\begin{equation}
\partial^{\dot bb}_\mu\vp^{\dot aa}_\nu(x)\longrightarrow \vp(x)^{-1}\vp^{\dot bb}_\mu(x)\vp^{\dot aa}_\nu(x)\label{eq:remap}
\end{equation}
with the sole exception $\dot a=\dot b$, $a=b$.\footnote{To confirm \pEref{eq:remap} for vector boson fields sharing a single \protect{$R$}-index, recognise that there always exists a global \protect{$\GL{\N}{R}$} transformation whose associated Lie algebra homomorphism maps any pair \protect{$\vp^{\dot aa}_\mu\vp^{\dot ab}_\nu$} or \protect{$\vp^{\dot aa}_\mu\vp^{\dot ba}_\nu$} onto a pair which does not share an \protect{$R$}-index, e.g.~\protect{$\vp^{\dot aa}_\mu\vp^{\dot bb}_\nu$}.} 
While operator $\bm{\partial}_\mu$ necessarily contains $\partial_\mu^{\dot aa}$ when acting on a field $\vp^{\dot aa}_\mu$, it also contains many other terms. 
It is seen in \sref{sec:initlagr} that the terms of $\bm{\partial}_\mu$ collectively act %
as a covariant derivative $D_\mu$, within which $\partial_\mu^{\dot aa}$ plays the role of the spacetime derivative when acting on $\vp^{\dot aa}_\mu$.
Therefore define
\begin{equation}
\partial_\mu\vp^{\dot aa}_\nu(x):=\partial^{\dot aa}_\mu\vp^{\dot aa}_\nu(x).\label{eq:genmudivvector}
\end{equation}
For the scalar boson, as previously, let the spatial derivative be defined through consistency with
\begin{equation}
\partial^\mu\partial_\mu \h(x):=-2\bar\partial_\rmU\partial_\rmU\h(x),\label{eq:genmudivscalar}
\end{equation}
and note that all $R$-indices are summed, so derivation of \Eref{eq:genmudivscalar} is unchanged from \cref{ch:simplest}.
Finally, for the spinor fields, hermeticity requires that the propagator (or equivalently, for massive particles, the mass matrix) 
be diagonalisable. %
Since
\begin{equation}
(\bar\partial^{\dot a}\bsm\partial^a)\psi^b_\alpha(x)=\vp(x)^{-1}[\bar\partial^{\dot a}\bsm\psi^a(x)]\psi^b_\alpha(x)\label{eq:genmudivspinor}
\end{equation}
for all $a$ except $a=b$, %
the spatial derivative must contain $\partial^a_\alpha$.
Hermeticity then implies that in a basis which diagonalises the propagator, $\dot a=a$.
In such a basis, it then follows that the only choice for $\partial_\mu\psi^a_\alpha$ is
\begin{equation}
\partial_\mu\psi^a_\alpha(x):=\left.\partial^{\dot aa}_\mu\psi^a_\alpha(x)\right|_{\dot a=a}.
\end{equation}
As noted above \Eref{eq:genmudivvector}, the other components of $\bm{\partial}_\mu$ in \Eref{eq:genmudiv} once again combine with this definition of $\partial_\mu$ to yield a covariant derivative, $D_\mu$. %

As in \sref{sec:fgspinorscalar}, %
linear independence permits any vector derivative operator other than those identified as the space--time derivatives to be replaced according to the rule
\begin{equation}
\partial^{\dot mn}_\mu=\bar\partial^{\dot m}\bsmm\partial^n\longrightarrow\vp(x)^{-1}\vp^{\dot mn}_\mu,
\end{equation}
with $\vp(x)^{-1}$ %
admitting replacement by $\la\vp(x)^{-1}\ra_{\mc{L}_0}$ in the presence of a pseudovacuum in the low-energy limit.

\subsection{Initial Lagrangian on \protect{$\RM$}\label{sec:initlagr}}

As in \sref{sec:themodel}, %
the Lagrangian on $\RM$ arises from the exchange statistics of the derivative operators on $\Cwtn$. Since the derivatives on $\RM$ are always constructed from spinor operators carrying the same $R$-indices as the objects on which they act, these constraints are identical to those observed in the previous chapter. If $\vp_q(x)$ is a single field on $\Cwtn$, with derivative fields 
\begin{align}
\psi^{a\alpha}_q=\partial^{a\alpha}\vp_q\qquad&\bar\psi_{q\dot a\dot\alpha}=\partial_{\dot a\dot\alpha}\vp_q\\
\h_q=\partial_\rmU\vp_q\qquad&\bar\h_q=\bar\partial_\rmU\vp_q\\
\vp_{q\mu}^{\dot aa}=\partial^{\dot aa}_\mu\vp_q,\qquad
\end{align}
these conditions are
\begin{align}
\rmi\partial_\mu\partial_\nu(\varphi_{q\rho})&=0\Ptagref{I}{eq:RMconstr1}\\
\partial_\mu(\partial_\nu\psi^\alpha_{q})&=0\Ptagref{I}{eq:RMconstr2}\\
\partial_\mu(\partial_\nu\bar\psi_{q\dot\alpha})&=0\Ptagref{I}{eq:RMconstr3}\\
\partial_\mu(\h^{\alpha\beta}_{q})&=0\Ptagref{I}{eq:RMconstr4}\\
\partial_\mu(\bar\h_{q\dot\alpha\dot\beta})&=0\Ptagref{I}{eq:RMconstr5}
\end{align}
on $\RM$.

Introduction of the pseudovacuum and foreground fields proceeds similarly to \srefr{sec:QLintro}{sec:quasi}, %
with the additional requirement that the pseudovacuum be \emph{on average} $\GL{\N}{R}$-symmetric over length and time scales large compared with the characteristic scale of the pseudovacuum, $\mc{L}_0$.
Construction of the expanded derivative operator and effective Lagrangian also proceeds equivalently%
. Writing the resulting derivative operator as
\begin{align}
{D}_\mu:=\partial_\mu&-\rmi fe_{\dot aa}\fgfield{\vp^{\dot aa}_\mu-\rmi f\bar\psi^{\dot a}\bsmm\psi^a}\label{eq:II:expandedD}
\\&-\frac{\rmi f}{2%
}\fgfield{\Upsilon_\mu(\h+\rmi f\psi^a\psi_a)+\bar\Upsilon_\mu(\h^*+\rmi f\bar\psi_{\dot a}\bar\psi^{\dot a})}\nn
\end{align}
where
\begin{equation}
f:=\la\vp^{-1}\ra_{\mc{L}_0}\approx\la\bgfield{\vp^{-1}}\ra_{\mc{L}_0}
\end{equation}
and defining
\begin{align}
F_{\mu\nu}&:={D}_\mu{D}_\nu 1-{D}_\nu{D}_\mu 1\label{eq:II:L1}\\
\tilde{D_\mu}&:=D_\mu+\frac{\rmi}{2} \tilde{\Upsilon}_{\mu}m_\psi\\
{{\triangle}}^{\mu\nu}&:=\eta^{\mu\nu}{{\square}}-{D}^\mu {D}^\nu\label{eq:II:ttr}\\
{{\square}}&:={D}^\mu {D}_{\mu}\label{eq:II:tsq},
\end{align}
the effective Lagrangian of the foreground fields takes the form
\begin{align}
\mscr{L}_\mrm{fg}=\Tr\Big\{&-\frac{1}{4}\fgfield{F^{\mu\nu}F_{\mu\nu}}+\rmi\fgfield{(\tilde{D}^\mu{\bar\psi})\,\tilde{\Dslash}(\tilde{D}_\mu{\psi})}\label{eq:II:L}\\
&-m^2_{\vp}e_{\dot aa}e_{\dot bb}\fgfield{\varphi^{\dot aa\mu}\vp^{\dot bb}_\mu}-m^2_\h\fgfield{\h^*\h}\Big\}\nn,
\end{align}
subject to the constraint that 
\begin{itemize}
\item where $\Upsilon_\mu\psi^a\psi_a$ and $\bar\Upsilon_\mu\bar\psi_{\dot a}\bar\psi^{\dot a}$ appear in the Lagrangian, these must be paired with $\bar\Upsilon_\nu\h^*$ and $\Upsilon_\nu\h$ respectively,
\end{itemize} 
and the recognition that
\begin{itemize}
\item the vector and scalar bosons are no more than shorthands for pairs of preons sharing a common source and sink, and so \emph{no two preon lines may have in common both terminating vertices} to avoid double counting. (For an example of this, see \sref{sec:fgspinorscalar}.) %
\end{itemize}
The equations of motion are\footnote{Derivation of the preonic equations of motion \perefr{eq:II:fgEOM2}{eq:II:fgEOM3} depends on emergent charge superselection rules, discussed in \pcref{ch:fermion}, %
which addresses the mass mechanism for fermions.} %
\begin{align}
{\triangle}^{\mu\nu}\fgfield{\vp^{\dot aa}_\nu}&=m^2_\vp\fgfield{\vp^{\dot aa\mu}}\label{eq:II:fgEOM1}\\
\Dslash\,\Dslash\,\fgfield{\psi^a_{\alpha}}&=m^2_\psi \fgfield{\psi^a_{\alpha}}\label{eq:II:fgEOM2}\\
\Dslash\,\Dslash\,\fgfield{\bar\psi^{\dot a\dot\alpha}}&=m^2_\psi\fgfield{\bar\psi^{\dot a\dot\alpha}}\label{eq:II:fgEOM3}\\
{\square}\fgfield{\h}=m^2_\h\fgfield{\h}\quad&\quad{\square}\fgfield{\h^*}=m^2_\h\fgfield{\h^*}.\label{eq:II:fgEOM5}
\end{align}

Note that the use of the covariant derivative on the bundles over $\RM$ automatically yields the matrix commutator term in the field strength tensor. That is, when $F_{\mu\nu}$ is defined as per \PEref{I}{eq:I:L1},
\begin{equation}
F_{\mu\nu}:=D_\mu D_\nu 1-D_\nu D_\mu 1,\label{eq:FDD1}
\end{equation}
and an abbreviation is introduced
\begin{equation}
\tilde{\vp}^{\dot aa}_\mu:=\vp^{\dot aa}_\mu-\rmi f\bar\psi^{\dot a}\bsmm\psi^a,\label{eq:tvp}
\end{equation}
then the vector boson terms in \Eref{eq:FDD1} evaluate as
\begin{equation}
\begin{split}
D_\mu &\tvp^{\dot aa}_{\nu}e_{\dot aa}-D_\nu \tvp^{\dot aa}_{\mu}e_{\dot aa}\\
&=\partial_\mu \tvp^{\dot aa}_{\nu} e_{\dot aa}-\partial_\nu \tvp^{\dot aa}_{\mu} e_{\dot aa} - \rmi f \left[\tvp^{\dot aa}_{\mu} e_{\dot aa},\tvp^{\dot bb}_{\nu} e_{\dot bb}\right]
\end{split}\label{eq:[]}
\end{equation}
as expected for a nonabelian gauge symmetry.

\subsection{Local symmetries of $\mscr{L}_\fg$\label{sec:localsym}}

It is now interesting to compare the structure of the derivative operator~\eref{eq:II:expandedD} with the decomposition of the global $\GL{2\N}{C}$ symmetry given in \Eref{eq:GL2NCdecomp2new}. Beginning with symmetry group $\SU{\N}$, the associated Lie algebra $\su{\N}$ admits a representation on $\N\times\N$ matrices comprising $\N-1$ diagonal matrices, and $\N^2-\N$ Hermitian matrices for which all diagonal elements are zero. The Lie algebra $\gl{1}{R}$ admits a representation on the $\N\times\N$ matrices comprising the single basis element $\frac{1}{\sqrt{\N}}\mbb{I}_\N$. Collectively, the $\N\times\N$ representation of Lie algebras $\su{\N}\oplus\gl{1}{R}$ spans the space of Hermitian matrices. As a concrete example, for $\N=3$, the off-diagonal matrices are given by 
\begin{align}
\begin{array}{cc}
\ulambda_1=\frac{1}{\sqrt{2}}\bgrid 0&1&0\\1&0&0\\0&0&0 \egrid &
\ulambda_2=\frac{1}{\sqrt{2}}\bgrid 0&-\rmi&0\\\rmi&0&0\\0&0&0 \egrid \\
\ulambda_4=\frac{1}{\sqrt{2}}\bgrid 0&0&1\\0&0&0\\1&0&0 \egrid &
\ulambda_5=\frac{1}{\sqrt{2}}\bgrid 0&0&-\rmi\\0&0&0\\\rmi&0&0 \egrid \\
\ulambda_6=\frac{1}{\sqrt{2}}\bgrid 0&0&0\\0&0&1\\0&1&0 \egrid &
\ulambda_7=\frac{1}{\sqrt{2}}\bgrid 0&0&0\\0&0&-\rmi\\0&\rmi&0 \egrid 
\end{array}\label{eq:Cbasis1}
\end{align}
where the numbering is chosen to match the off-diagonal Gell-Mann matrices $\{\bm{\lambda}_\ta\}$, $\ulambda_\ta=\bm{\lambda}_\ta/\sqrt{2}$. Indices carry a tilde ($\ta$) to show that in $\Cwtn$ models they are associated with neither the holomorphic nor the antiholomorphic sector. The corresponding diagonal matrices are given by rescaling the remaining two Gell-Mann matrices and the identity,
\begin{align}
\begin{array}{cc}
\ulambda_3=\frac{1}{\sqrt{2}}\bgrid 1&0&0\\0&-1&0\\0&0&0 \egrid &
\ulambda_8=\frac{2}{\sqrt{6}}\bgrid \frac{1}{2}&0&0\\0&\frac{1}{2}&0\\0&0&-1 \egrid\\
\ulambda_9=\frac{1}{\sqrt{3}}\bgrid 1&0&0\\0&1&0\\0&0&1 \egrid.
\end{array}\label{eq:Cbasis2}
\end{align}
In a model which exhibits a local $\SU{\N}\oplus\GL{1}{R}$ symmetry, each such matrix $\ulambda_\ta$ will be associated with a vector boson $\vp^\ta_\mu$. Linear recombination of these bosons permits re-expression in terms of $(\N^2-\N)/2$ complex vector bosons associated with the matrices $e_{\dot aa}|_{\dot a<a}$ and $\N$ real vector bosons associated with the matrices $e_{\dot aa}$. For example, on $\SU{3}$ define
\begin{equation}
\begin{split}
\vp^{12}_\mu &= \frac{1}{\sqrt{2}}\left(\vp^1_\mu+\rmi\vp^2_\mu\right)\qquad\vp^{21}_\mu=\vp^{12\dagger}_\mu
\\
\vp^{13}_\mu &= \frac{1}{\sqrt{2}}\left(\vp^4_\mu+\rmi\vp^5_\mu\right)\qquad\vp^{31}_\mu=\vp^{13\dagger}_\mu\\
\vp^{23}_\mu &= \frac{1}{\sqrt{2}}\left(\vp^6_\mu+\rmi\vp^7_\mu\right)\qquad\vp^{23}_\mu=\vp^{32\dagger}_\mu\\
\vp^{11}_\mu &= \frac{1}{\sqrt{2}}\vp^3_\mu+\frac{1}{\sqrt{6}}\vp^8_\mu+\frac{1}{\sqrt{3}}\vp^9_\mu\\
\vp^{22}_\mu &= -\frac{1}{\sqrt{2}}\vp^3_\mu+\frac{1}{\sqrt{6}}\vp^8_\mu+\frac{1}{\sqrt{3}}\vp^9_\mu\\
\vp^{33}_\mu &= -\frac{2}{\sqrt{6}}\vp^8_\mu+\frac{1}{\sqrt{3}}\vp^9_\mu.
\end{split}
\end{equation}
This is precisely the symmetry structure seen in the vector boson component of \Eref{eq:II:expandedD}, though with
the Hermitian matrix-valued gauge field $e_{\dot aa}\vp^{\dot aa}_\mu$ replaced by 
\begin{equation}
e_{\dot aa}(\vp^{\dot aa}_\mu-\rmi f\bar\psi^{\dot a}\bsmm\psi^a), \label{eq:compositegaugefield}
\end{equation}
reflecting the preonic substructure of the emergent vector bosons.
As in \sref{sec:fgspinorscalar}, it is implicit that the preonic decomposition is only used in contexts where no two preons share a common source or sink, and thus $\vp^{\dot aa}_\mu$ may be seen as a sometimes-convenient restatement of $-\rmi f\bar\psi^{\dot a}\bsmm\psi^a$. In models having energy scales at which the preon substructure may be ignored (see \sref{sec:bosonsinn=3}), \Eref{eq:compositegaugefield} reduces to a conventional gauge field $e_{\dot aa}\vp^{\dot aa}_\mu$ for $\su{\N}\oplus\gl{1}{R}$.
Furthermore, recognising that the resulting gauge bosons are associated with the $[\SU{\N}\oplus\GL{1}{R}]\otimes\SL{2}{C}$ component of \Eref{eq:GL2NCdecomp2new}, observe that the vector character of the bosons arises from the $\SL{2}{C}$ symmetry of the cotangent bundle over $\RM$. On the aforementioned models where the preon components may be ignored, these gauge bosons are therefore associated with changes in co-ordinates on an implied $\SU{\N}\oplus\GL{1}{R}$-valued {cotangent} bundle over $\RM$, i.e.~$[\SU{\N}\oplus\GL{1}{R}]\otimes T^*\RM$.

The second part of \Eref{eq:II:expandedD} admits a similar explanation, in which the complex scalar boson $\h$ and its preonic decomposition are associated with the symmetry $[\SU{\N}\oplus\GL{1}{R}]\otimes\mbb{C}^\times$. %
This term also admits a geometric description, this time corresponding to changes in co-ordinates on
a $\mbb{C}^\times$ bundle over $\RM$. The emergent representation of $\SU{\N}\oplus\GL{1}{R}$ %
which is associated with this bundle is trivial (dimension~1), and consequently the representation of the Lie algebra on the $\SU{\N}\oplus\GL{1}{R}$-valued $\mbb{C}^\times$-bundle over $\RM$ is two-dimensional, with basis corresponding to the real and imaginary parts of $\h(x)$, or equivalently to $\h(x)$ and $\h^*(x)$.

Pulling all of this together, it follows that in contexts where preonic decompositions may be ignored, the derivative operator given in \Eref{eq:II:expandedD} may be understood as a covariant derivative with respect to $\SU{\N}\oplus\GL{1}{R}$-valued cotangent and $\mbb{C}^\times$ bundles over $\RM$. The form of $\mscr{L}_\fg$~\eref{eq:II:L} is then seen to be the natural Lagrangian of the emergent locally symmetric gauge theory, up to the mass terms arising from the pseudovacuum which break the $\mbb{R}^+$ %
component of the 
global $\mbb{C}^\times$ symmetry. %
In \sref{sec:initlagr} it was assumed that the $\GL{\N}{R}$ symmetry which gives rise to $\SU{\N}\oplus\GL{1}{R}$ remains unbroken, and thus all vector bosons and preons have identical masses $m_\vp$ and $m_\psi$ respectively. This assumption is revisited in \sref{sec:gauge}.

Recognising that the local symmetry of the foreground fields is an emergent phenomenon, it is appropriate to seek out its origins on %
$\Cwtn$. %
This local freedom of basis on $\SU{\N}\oplus\GL{1}{R}$ corresponds to a freedom to adopt co-ordinate frames on $\Cwtn$ with nonvanishing connection.\footnote{On adopting a co-ordinate frame on \prm{\Cwtn} with non-vanishing connection, the construction of fields and operators on \prm{\RM} is then with respect to the covariant derivative \prm{\nabla^a_\alpha} rather than \prm{\partial^a_\alpha}. Thus, for example, \Eref{eq:vpaa} becomes
\begin{equation*}
\left.\vp^{\dot aa}_\mu(x)\,\right|_\RM:=\left.\partial^{\dot aa}_\mu\,\vp[\G^{-1}(x)]\,\right|_\Cwtn\qquad\partial^{\dot aa}_\mu:=\bar\nabla^{\dot a}\bsmm\nabla^a. \tag{\ref{eq:vpaa}a}
\end{equation*}}
Local redefinitions of the $\SU{\N}\oplus\GL{1}{R}$ gauge fields are then associated with local $\SU{\N}\oplus\GL{1}{R}$ mixing, acting on the $R$-indices of the co-ordinate frame, while changes of co-ordinates on $\SL{2}{C}$ or on the $\mbb{C}^\times$ bundle are realised through the action of unitaries on the $S$-indices of the co-ordinate frame on $\Cwtn$. %
It should be noted that nonvanishing connections on either $R$ or $S$ indices do not necessarily require nonvanishing values for co-ordinate--independent measures of curvature. A close analogue to this mixing is the Coriolis effect in a rotating frame of reference, which also introduces connections without affecting frame-independent measures of curvature.

As an aside, note that all derivative operators appearing in Lagrangian~\eref{eq:II:L} are with respect to $\RM$, with the multiplicity of directions on the $\Cwtn$ bulk mapping onto multiple fields on $\RM$. The lack of explicit off-$\RM$ directions appearing in $\mscr{L}$ justifies the decision to study a quasiparticle model on a $\Cwt$ subspace of $\Cwtn$ rather than on $\Cwtn$ entire, as the low-energy dynamics of the derivative fields \erefr{eq:Cwtnfields1}{eq:Cwtnfields5} on $\RM$ may be extrapolated without reference to their values at locations on the $\Cwtn$ bulk extraneous to $\Cw{2\bullet}$.

\section{Colour interaction}

\subsection{Binding preons into bosons in \protect{$\N=3$}\label{sec:bosonsinn=3}}

Specialising now to $\N=3$ and adopting basis~\erefr{eq:Cbasis1}{eq:Cbasis2}, it follows from \sref{sec:localsym} that the vector bosons and their associated preonic re-expressions %
in derivative~\eref{eq:II:expandedD} mediate an $\SU{3}$ gauge field interaction.
It is widely accepted that the presence of an $\SU{3}$ gauge field implies confinement of species carrying charge with respect to this field (i.e.~confinement of ``coloured'' particle species). In the $\Cws$ model, all preons carry charges with respect to a nontrivial representation of $\SU{3}$, and are therefore confined.
Let the energy scale of %
this preon confinement be denoted $\mc{E}_\preon$, and associated with a confinement length scale $\mc{L}_\preon$. Calculation of the confinement scale is beyond the scope of this chapter, so for the present it is assumed to be equal to the energy scale at which the emergent quasiparticle model diverges from the emulated QFT. Composite bosons are therefore assumed tightly bound at all energy scales considered in the current chapter. This divergence scale is %
shown in \Psref{I}{sec:pushlimits} %
to be 
\begin{equation}
\frac{1}{2}\mc{E}_\Omega:=\frac{\N}{2} N_0\mc{E}_0.
\end{equation}

Regarding the binding of preons, two broad categories of colour-neutral grouping have already been identified: the preon pairings in diagonal vector bosons, and the pairing associated with the scalar boson. 
As a matter of notation, let the $R$-index values $\{1,2,3\}$ be replaced by the colours $\{r,g,b\}$ or the anti-colours $\{\dot r,\dot g,\dot b\}$ on the holomorphic or antiholomorphic sectors respectively.
The diagonal vector bosons are then colour-neutral term by term, e.g.{}
\begin{equation}
\vp^3_\mu\longleftrightarrow\frac{f}{\sqrt{2}}\left(\bar\psi^{\dot r}\bsmm\psi^r-\bar\psi^{\dot g}\bsmm\psi^g\right).\label{eq:vplr}
\end{equation}
The local $\SU{3}$ symmetry therefore binds the preonic constituents of the diagonal bosons to prevent exposure of their colour charges. It then follows that this binding %
also applies
to the constituents of the off-diagonal vector bosons by the unbroken nature of the $\SU{3}$ symmetry.

In contrast with the vector bosons, which are neutral term by term, the scalar boson decomposes into a sum over preon pairs which are individually coloured, but collectively colourless as the sum over the $R$-index renders trivial the action of $\SU{3}$ on the sum as a whole:
\begin{equation}
\h\longleftrightarrow f\!\!\sum_{c\in\{r,g,b\}}\!\!\psi^{c}\psi_c.\label{eq:Hlr}
\end{equation}
The need to maintain colour neutrality therefore also serves to bind together the preonic constituents of the scalar boson and prevent their individual coupling to the colour indices of other (probe) species.

Given the existence of an energy scale $\mc{E}_\preon$ associated with the confinement of preons, it then follows that free preons are not observed at energy scales small compared with $\mc{E}_\preon$, and the derivative operator~\eref{eq:II:expandedD} reduces to
\begin{align}
D_\mu=\partial_\mu&-\rmi \bmmf\ulambda_\ta\fgfield{\bmvp^{\ta}_\mu}
-\frac{\rmi \bmmf'}{2}\fgfield{\Upsilon_\mu\bmh+\bar\Upsilon_\mu\bmh^*},\label{eq:interimD}
\end{align}
consistent with the geometric interpretation introduced in \sref{sec:localsym}. 

In going from %
\Eref{eq:II:expandedD} to \Eref{eq:interimD} a subtle but important change in the boson fields has taken place, reflected by the bolding of the fields and coupling coefficients. In the boson fields of \Eref{eq:II:expandedD}, two chiral derivative operators acted on the product field $\vp(x)$ at the same point. However, in \Eref{eq:interimD} the boson fields are defined as corresponding to a pair of chiral derivative operators confined by the colour interaction, acting at points separated by not more than $\ILO{\mc{L}_\preon}$. While this description includes the field configurations of the previous definition, it also includes states in which the boson separates into a non-collocated pair of preons, provided the separation of these preons is negligible in the low-energy regime. 

There are multiplicative numerical factors associated with this change in definition. First there is the factor $f$ seen in \Erefr{eq:vplr}{eq:Hlr}, corresponding to $\la\vp\ra^{-1}$, as a vector or scalar boson contains one copy of $\vp$~(\notchap{I:}\ref{eq:defvarphimu},\notchap{I:}\ref{eq:defHproduct}) but a pair of preons contains two~\Peref{I}{eq:defspinors}. However, these are not the only factors.
In \sref{sec:bgfieldsequiv}, the background field correlator 
$\la\bgfield{\vp^\mu\vp_\mu}\ra_{\mc{L}_0}$%
was shown to evaluate to ${N_0}^2{\omega_0}^2$. In the first step of this calculation the substitution of \PEref{I}{eq:vpsub0} into the correlator %
yielded ${N_0}^4$ terms, each containing four chiral derivatives. Properties of the background fields then forced all of these derivative operators to act on the same fundamental scalar field, reducing the number of terms to ${N_0}^2$. Foreground fields, however, do not satisfy the properties of the background fields used to achieve this reduction. Instead, in general, they attract a factor of $N_0$ for each chiral derivative in their expansion. This factor is achieved by reordering the fundamental scalar fields, which have bosonic symmetry, and not the chiral derivatives, which are fermionic, and there are consequently no factors of $-1$ to generate destructive interference. This holds even when the foreground species attracting the multiplicative factor is fermionic. %

The extra factors of $N_0$ can in general be absorbed into a redefinition of the coupling constant $f$, though the details of this redefinition may depend on the species participating in an interaction. Explicit examples are evaluated in \cref{ch:boson}, e.g.~\sref{sec:vecbosonmasses}. %
In the present chapter the $\SU{3}\oplus\GL{1}{R}$ symmetry is assumed unbroken, such that all bosons $\fgfield{\bmvp^\ta_\mu}$ in \Eref{eq:interimD} have been assigned identical coupling coefficients $\bmmf$. However, in the interest of retaining some generality a distinct coefficient $\bmmf'$ has been introduced for the scalar field $\bmh$.

In principle the coloured vector bosons (gluons) continue to appear in the derivative $D_\mu$ as shown in \Eref{eq:interimD}, but these particles are only observed in the presence of coloured sources. Thus in a model with multiple coloured species having confinement energy scales $\mc{E}_\preon$, $\mc{E}_{\preon'}$, etc., gluons display effective confinement with an energy scale of at least $\min{\{\mc{E}_\preon,\mc{E}_{\preon'},\ldots\}}$. %
Free gluons may therefore also be suppressed at energy scales $\mc{E}\ll\mc{E}_\preon$, and \Eref{eq:interimD} reduces to
\begin{align}
\bm{D}_\mu:=\partial_\mu&-\frac{\rmi \bmmf}{\sqrt{3}}\fgfield{\bmvp^{9}_\mu}\label{eq:reducedD}
-\frac{\rmi \bmmf'}{2}\fgfield{\Upsilon_\mu\bmh+\bar\Upsilon_\mu\bmh^*}.
\end{align}
It is interesting to compare this expression with derivative operator~\Peref{I}{eq:Dbold} in which the preon terms must remain extant, as there does not exist a mechanism of confinement at $\N=1$ to guarantee their disappearance in the low-energy limit. Derivative operators~\erefs{eq:interimD}{eq:reducedD} are more concise, but are only valid at energies small compared with the preon binding scale. At higher energies it remains necessary to articulate the $\SU{3}$ symmetry responsible for confinement using both fermions and preons, but defining
\begin{equation}
\tilde{\vp}^{\dot aa}_\mu:=\vp^{\dot aa}_\mu-\rmi f\bar\psi^{\dot a}\bsmm\psi^a,\tag{\ref{eq:tvp}}
\end{equation}
it is easy to see from \Eref{eq:[]} that this local symmetry remains extant even on approaching the preon scale.

Having observed that there exists an energy scale at which all preons are bound, it is now desirable to look for any other colourless composites which may exist in $\Cws$. As it happens, another family of colourless composite particles does exist, being fermions with spin $\frac{1}{2}$ and units $L^{-3/2}$. These species act as fermionic matter fields in the low energy limit.

\subsection{Composite fermions in \protect{$\N=3$}\label{sec:compfermi}}

The next colour-neutral construction to consider is a triplet of collocated preons, one of each colour, which enjoys net colour invariance under the action of $\SU{3}$. Such a construction may involve two internally summed indices, e.g.{}
\begin{equation}
\psi^{r\alpha}\psi^g_\alpha\psi^b_\beta,\label{eq:fermionexample}
\end{equation}
and indeed, at energy scales $\mc{E}\ll \mc{E}_\preon$ at which there are no free preons, the only Lagrangian terms involving neutral triplets which can be constructed on $\RM$ are those in which this internal summation does take place. The preons in this construction may be truly collocated, or they may be separated by length scales up to $\ILO{\mc{L}_\preon}$, with the distinction being irrelevant at energies $\mc{E}\ll \mc{E}_\preon$.

Given such a triplet, note that summation over an $S$-index makes it %
invariant under the action of $\SU{3}$ only up to cyclic permutation of colour indices. (That is, for example, $\psi^{r\alpha}\psi^g_\alpha\psi^b_\beta$ is distinct from $\psi^{g\alpha}\psi^b_\alpha\psi^r_\beta$ etc.)
In contrast, the Lagrangian as a whole \emph{is} invariant under $\SU{3}$ global transformations, which includes invariance under the discrete group of cyclic permutations of colour labels. Taken together with hermeticity, colour cycle invariance suffices to imply that in the limit of a large number of interactions between preons and $\su{3}$ vector bosons, the action of the $\SU{3}$ sector on the space of colour charges takes the general form \cite{koide2000}
\begin{align}
\label{eq:SU3action}
\left(\begin{array}{c}\psi^{r'}\\\psi^{g'}\\\psi^{b'}\end{array}\right) &= K \left(\begin{array}{c}\psi^{r}\\\psi^{g}\\\psi^{b}\end{array}\right)\\
K = \frac{a_\psi}{\sqrt{3}} \mbb{I}_3 \,-\, &\frac{b_\psi}{\sqrt{6}} \bgrid 0 & e^{\rmi\thps} & e^{-\rmi\thps}\\
e^{-\rmi\thps} & 0 & e^{\rmi\thps}\\
e^{\rmi\thps} & e^{-\rmi\thps} & 0 \egrid\label{eq:II:KfES}
\end{align}
where parameters $a_\psi$, $b_\psi$, and $\thps$ are constants which are fixed by the model.
While the factors of $1/\sqrt{3}$ and $-1/\sqrt{6}$ could be absorbed into $a_\psi$ and $b_\psi$ respectively, they are left extant by convention so that \Eref{eq:II:KfES} corresponds to the form of the Koide matrix $K$ in \rcite{koide2000}.
In the $\Cwtn$ series, matrices of this form appear for all models in subseries $\Cw{6\N}$, and the values of parameters $a_\psi$, $b_\psi$, and $\thps$ can be calculated from first principles. However, $\thps$ in particular is both model- and species-dependent and for now it suffices to note that matrix $K$ is diagonalisable, and its eigenvalues are positive.
Positivity of eigenvalues follows from requiring that a full cyclic permutation of colours ($r\rightarrow g\rightarrow b\rightarrow r$ and similar), which is in $K^3$, %
must leave the sign of an eigenstate of $K$ unchanged in accordance with the colour cycle invariance of the unbroken $\SU{3}$ symmetry.

Recognising that the preons in the triplet are not well-localised, detection of a preon at any given point may yield a red, green, or blue preon with differing probabilities. In this context, the column vectors in \Eref{eq:SU3action} may represent the amplitudes of the different colour preon fields at a specific point $P$. However, it is more useful to assume integration over a spatial volume of $\ILO{{\mc{L}_\preon}^3}$ in the net rest frame of the composite fermion, in which context \Eref{eq:SU3action} represents exchange of the different preon colours in the triplet over multiple interactions mediated by $\su{3}$ [or, more precisely, $\su{3}\oplus\gl{1}{R}$] vector bosons.
Diagonalisation of the mixing matrix then yields three orthogonal eigenvectors, corresponding to triplets which remain on average invariant under sustained $\su{3}$ vector boson exchange and thus behave as if they are, collectively, colourless. These eigenvectors may be written $\Psi^{g}_\alpha$ where $g$ enumerates the eigenvectors, and may be considered the ``generation'' of fermion triplet $\Psi^g_\alpha$. This identification is made more formal in \cref{ch:fermion}. %

\subsection{Lagrangian without preons}

In the limit $\mc{E}\ll \mc{E}_\preon$, all preons are bound into colourless composite objects when examined over length scales $\mc{L}\gg \mc{L}_\preon$. 
A species is colourless either if it carries no $R$-indices, and is thus associated with the trivial representation of $\SU{3}$, or equivalently, if it is explicitly associated with the identity matrix, also corresponding to the trivial representation of $\SU{3}$. 
In the boson sector, the foreground boson fields $\bmvp^9_\mu$, $\bmh$, and $\bmh^*$ are therefore colourless and unconfined.
Through binding of fermion triplets (\sref{sec:compfermi}), the model has also gained a new family of colourless fermions denoted $\Psi^g_\alpha$ and their conjugates. In the low-energy limit the derivative operator therefore reduces to
\begin{align}
\bm{D}_\mu:=\partial_\mu&-\frac{\rmi \bmmf}{\sqrt{3}}\fgfield{\bmvp^{9}_\mu}\tag{\ref{eq:reducedD}}
-\frac{\rmi \bmmf'}{2}\fgfield{\Upsilon_\mu\bmh+\bar\Upsilon_\mu\bmh^*},
\end{align}
as indicated above, with field strength tensor
\begin{equation}
\bm{F}_{\mu\nu}:=\bm{D}_\mu\bm{D}_\nu 1-\bm{D}_\nu\bm{D}_\mu 1,
\end{equation}
and with appropriate normalisation of $\Psi^g_\alpha$ the Lagrangian becomes
\begin{align}
\begin{split}
\!\!\!\mscr{L}_\mrm{fg}=&-\frac{1}{4}\bm{F}^{\mu\nu}\bm{F}_{\mu\nu}+\rmi\fgfield{\bar\Psi^g\,{\bmDslash}\Psi_g}-m^2_{\bmvp}\fgfield{\bmvp^{9\mu}\bmvp^{9}_\mu}\\
&-m_\Psi\fgfield{\Psi^g\Psi_g}-m_\Psi\fgfield{\bar\Psi^g\bar\Psi_g}-m^2_\bmh\fgfield{\bmh^*\bmh}.%
\end{split}\label{eq:L2}
\end{align}
This Lagrangian exhibits three fermions and three antifermions of dimension $L^{-3/2}$ and spin\ $\frac{1}{2}$, interacting by exchange of one species of vector boson and one species of complex scalar boson.
The fermion term arises naturally from the action of the covariant derivative $\bm{D}_\mu$ on the constituent preons of $\Psi_g$, much as discussed in \sref{sec:fgspinorscalar}.

\subsection{Gauge\label{sec:gauge}}

Following the construction described in \sref{sec:II:mapping}, the $\Cws$ model has an intrinsic symmetry \eref{eq:GL2NCdecomp2new} which is global on $\RM$ and may be expressed as 
\begin{equation}
\begin{split}
&\SU{3}\otimes\SL{2}{C}\\
\oplus\;&\GL{1}{R} \otimes\SL{2}{C}\\
\oplus\;&\SU{3}\otimes\mbb{C}^\times\\
\oplus\;&\GL{1}{R} \otimes\mbb{C}^\times.
\end{split}
\end{equation}
It also has a large number of composite boson fields which appear in the derivative operator~%
\eref{eq:interimD} %
and may be interpreted as %
gauge bosons, promoting elements of this symmetry from global on $\RM$ to local on $\RM$. Specifically, 
\begin{itemize}
\item[1.] the vector bosons $\fgfield{\bmvp^\ta_\mu}$, $\ta\in\{1,\ldots,8\}$ 
correspond to a dimension-8 representation of $\su{3}$ taking values on the cotangent bundle whose spinor form is $\SL{2}{C}$,
\item[2.] the vector boson $\fgfield{\bmvp^9_\mu}$ corresponds to a dimension-1 representation of $\gl{1}{R}$ taking values on the cotangent bundle whose spinor form is $\SL{2}{C}$.
\end{itemize}
In addition, 
\begin{itemize}
\item[3.] the scalar boson $\fgfield{\bmh}$ corresponds to a dimension-1 representation of $\su{3}\oplus\gl{1}{R}$ taking values on the complex bundle $\mbb{C}^\times$, and
\item[4.] the space--time connection %
is associated with a dimension-1 representation of $\su{3}\oplus\gl{1}{R}$. %
\end{itemize}
The space--time connection vanishes in Cartesian co-ordinates because mapping $\G$ is from $M\in\Cws$ to $\RM$. However, in other co-ordinates, or if the target manifold is replaced by a curved space--time with the same signature as the Minkowski metric, a connection $(\omega_\mu)^{\p{\beta}\alpha}_\beta$ %
may be realised on the $\SL{2}{C}$ representation of the cotangent bundle, with corresponding Christoffel symbols satisfying
\begin{equation}
(\omega_\mu)^{\p{\beta}\alpha}_\beta=\Gamma^\nu_{\mu\rho}\sigma_{\beta\dot\gamma\nu}\bar\sigma^{\dot\gamma\alpha\rho}.\label{eq:connectcoeff}
\end{equation}
The association with a dimension-1 representation of $\su{3}\oplus\gl{1}{R}$ as per item~4, above, indicates that this change in target manifold has no implications for choice of co-ordinates on the gauge group $\SU{3}\oplus\GL{1}{R}$. Curved target manifolds for mapping $\G$ are considered further in \cref{ch:gravity}. %

Similarly, item~3 associates a trivial representation of $\su{3}\oplus\gl{1}{R}$ with the connection on $\mbb{C}^\times$, with this connection being the complex boson $\fgfield{\bmh}$. As the representation of $\su{3}\oplus\gl{1}{R}$ is trivial, the complex boson carries no $R$-index. In principle there %
exist two gauge freedoms associated with arbitrary choice of connection on the $\mbb{C}^\times$-bundle. However, using Lemma~\eref{Lem:8} %
to rewrite
\begin{equation}
\bm{1}\otimes\mbb{C}^\times\cong\bm{1}\otimes\mbb{R}^+\oplus\bm{1}\otimes\U{1}
\end{equation}
where $\bm{1}$ is the trivial representation of $\SU{3}\oplus\GL{1}{R}$,
the introduction of the pseudovacuum characterised by an energy scale $\mc{E}_0$ breaks scale invariance, fixing the $\mbb{R}^+$ component of the $\mbb{C}^\times$ connection. The $\U{1}$ component remains unbroken, and thus there exists a gaugeable local $\U{1}$ symmetry mediated by the scalar boson field $\fgfield{\h}$. This gaugeable degree of freedom plays an important role in the model on $\Cw{18}$ (see \cref{ch:SM}). %
As per %
\sref{sec:localsym}, this gauging of the emergent local symmetry is realised on the underlying $\Cwtn$ manifold as corresponding to the adoption of an appropriate co-ordinate frame. %

For item~1, consider the $\su{3}$-valued vector boson field $\lambda_{\ta}\fgfield{\bmvp^{\ta}_\mu}$. This is the gauge field of the $\SU{3}\otimes\SL{2}{C}$ bundle, where $\SL{2}{C}$ is the spinorisation of the tangent bundle $T^*\RM$. The $\SU{3}$ component is gaugeable, while the $\SL{2}{C}$ component is fixed due to identification with the tangent bundle (which has a connection determined by curvature of the space--time manifold).
As per %
\sref{sec:localsym}, gauging of the emergent local $\SU{3}$ symmetry is realised on $\Cws$ by a change of co-ordinate frame, this time corresponding to the action of a position-dependent unitary matrix specifically on the $R$-indices of the original connection-free co-ordinate system.

The group $\SU{3}$ has five gaugeable degrees of freedom, as demonstrated in \aref{apdx:SU3gauge}.
These degrees of freedom may be used to impose constraints which break colour symmetry, either on the boson sector, e.g.{}
\begin{equation}
\fgfield{\bmvp^{1\mu}\bmvp^1_\mu}=0,\qquad\fgfield{\bmvp^{2\mu}\bmvp^3_\mu}=0,\qquad\mrm{etc.},\label{eq:gaugebosonsector}
\end{equation}
or on the fermion sector, e.g.{}
\begin{equation}
\psi^{r\alpha}\psi^r_\alpha=0,\qquad\psi^{g\alpha}\psi^b_\alpha=0,\qquad\mrm{etc.},\label{eq:gaugefermionsector}
\end{equation}
or by reference to some external variable presumed otherwise uncorrelated with colour, e.g.{}
\begin{equation}
\fgfield{\bmvp^{1\mu}\bmvp^1_\mu}=V(x),\qquad\mrm{etc.}\label{eq:externalgauge}
\end{equation}
Caution must be employed when introducing a choice of gauge, as this may either conceal or even break the internal $\SU{3}$ symmetry of the model. For example, choosing
\begin{equation}
\sum_{a\in\{r,g,b\}}\fgfield{\bmvp^{a\mu}\bmvp_{a\mu}}=\fgfield{\bmvp^{1\mu}\bmvp_{1\mu}}
\end{equation}
introduces a privileged relationship between the pseudovacuum and the boson associated with matrix $\ulambda_1$, breaking $\SU{3}$ symmetry, and thus may potentially yield a description in which confinement is not easily recognised. Such choices may also potentially break colour-independence of any properties of the model, such as the colour-independence of mass assumed in \Eref{eq:II:L}.

For all models in series $\Cw{6\N}$, it is possible to show the existence of an $\SU{3}\oplus\GL{1}{R}$ symmetry  through factorisation of $\GL{\N}{R}$ in \Eref{eq:GL2NCdecomp1}. For $\Cw{6\N}|_{\N>1}$ it is possible to construct internal gauge choices which do not necessarily break this symmetry. %
\Cref{ch:SM} considers gauge choices on $\Cw{18}$ in some detail, and addresses explicitly the construction of a choice of gauge which does not break the symmetry of the $\SU{3}$ colour subgroup of that model.

Finally, for item~2, the vector boson $\fgfield{\bmvp^9_\mu}$ is the gauge field of the $\GL{1}{R}\otimes\SL{2}{C}$ bundle, and by local isomorphism of $\GL{1}{R}$ and $\U{1}$ this field mediates a one-dimensional choice of gauge associated with a freedom of phase. Different choices of gauge may be preferred in different contexts and for different values of $\N$ in $\Cw{6\N}$, and again the choice of this gauge is discussed explicitly for $\Cw{18}$ in \cref{ch:SM}. %

The following observations from forthcoming chapters (\ref{ch:SM}--\ref{ch:boson}) %
are worth pre-empting, becoming
apparent when the mass mechanisms of the $\Cwtn$ model series are subsequently elaborated:
\begin{itemize}
\item Symmetry breaking on the fermion sector may lead to species-dependence of mass in the boson sector. This is the primary reason for deferring exploration of the mass mechanism, as the full details of the mass interactions depend explicitly on the value of $\N$ and the associated choices of gauge.
\item However, it is universally true that the coloured boson (gluon) masses are only observable at energy scales small compared with $\mc{E}_\Omega$. As $\mc{E}_\preon$ is taken to be of $\OO{\mc{E}_\Omega}$, %
the coloured bosons are massless at the energy scales which are associated with colour confinement. 
The $\SU{3}$ symmetry of the coloured boson masses therefore always remains effectively unbroken at these scales.
\item Since the bosons of the $\SU{3}$ symmetry in the $\Cws$ model are themselves coloured, they are confined. Their masses are never observed.
\item This is consistent with the treatment of gluons in Standard Model, in which the gluon mass is assumed to be zero.
\end{itemize}

\section{Conclusion}

In this chapter the $\Cwt$ model of \cref{ch:simplest} %
has been extended to $\Cwtn$, and the resulting models are seen to exhibit a vector boson sector corresponding to a local $\SU{\N}\oplus\GL{1}{R}$ symmetry. For $\Cws$ this includes an $\SU{3}$ colour interaction which binds together composite fermions with physically desirable properties, having dimension $L^{-3/2}$ and spin~$\frac{1}{2}$. In the low-energy limit these emergent composite fermions obey a simplified Lagrangian, interacting by exchange of one species of vector boson and one species of complex scalar boson. Once again this model is obtained entirely from the exchange statistics of vectors on the underlying manifold, and the introduction of a minimal-complexity %
pseudovacuum.

This result demonstrates that models in the $\Cwtn$ series are capable of supporting fermionic excitations with dimensions and spin consistent with the observed leptons (though having an additional composite substructure), while also providing a new toy model exhibiting a form of colour-mediated confinement. Subsequent extension of this work to higher values of $\N$ is capable of yielding dramatic increase in the richness of both the lepton and the boson sectors.

Subsequent chapters focus on the $\Cwtn$ model at $\N=9$, which contains particle species bearing
substantial resemblance to those of the Standard Model.
The mass mechanisms of the $\Cwtn$ series are described explicitly for this model, including the effects of choice of gauge.

\appendix

\section{Group identities\label{apdx:lemmas}}

The following definitions and group identities are used in this paper. %

\begin{lemma}[\ref{Lem:4}]
\begin{equation}
\GL{n}{C}\cong\SL{n}{C}\oplus\mbb{C}^\times\label{Lem:4}
\end{equation}
where $\mbb{C}^\times$ is the multiplicative group of complex numbers, also denoted $\GL{1}{C}$. %
\end{lemma}
\begin{lproof}
$\SL{n}{C}$ is $\GL{n}{C}$ modulo the determinant, which takes its value in $\mbb{C}^\times$.\qed\medskip\\
Note that the multiplicative group of real numbers may similarly also be denoted either $\mbb{R}^\times$ or $\GL{1}{R}$.
\end{lproof}

\begin{lemma}[\ref{Lem:8}]
\begin{equation}
\mbb{C}^\times\cong\mbb{R}^+\oplus\U{1}\label{Lem:8}
\end{equation}
where $\mbb{R}^+$, also denoted $\GLp{1}{R}$, is the identity component of $\mbb{R}^\times$ (i.e.~the connected component containing the identity).
\end{lemma}
\begin{lproof}
This is the decomposition of a complex number into magnitude and phase. \qed\medskip\\
Note that under appropriate logarithmic mappings, both $\mbb{R}^+$ and $\U{1}$ are isomorphic to $\mbb{R}$ up to the point on the boundary of $\mbb{R}^+$, which is only approached asymptotically, and an arbitrarily chosen point on $\U{1}$ which corresponds to the asymptote of $\mbb{R}$ at $\pm\infty$.
\end{lproof}

\begin{lemma}[\ref{Lem:11}]
\begin{equation}
\GL{n}{C}\cong\SL{n}{C}\oplus\mbb{R}^+\oplus\U{1},\label{Lem:11}
\end{equation}
as presented previously as \PEref{I}{eq:I:GL2Cdecomp}.
\end{lemma}
\begin{lproof}
Consecutive application of Lemmas~\erefs{Lem:4}{Lem:8}.\qed
\end{lproof}

\begin{definition}[\ref{Def:1}]
Let the complexification of a general Lie group $\mrm{G}$ be denoted $\mrm{G}(\mbb{C})$ and defined as %
\begin{equation}
\mrm{G}(\mbb{C}):=\mbb{C}\otimes_\mbb{R}\mrm{G}.\label{Def:1}
\end{equation}
For example,
\begin{equation*}
\begin{split}
\GL{a}{C}&\;\cong\mbb{C}\otimes_\mbb{R}\GL{a}{R}\\
\SU{n,\mbb{C}}&:=\mbb{C}\otimes_\mbb{R}\SU{n}.
\end{split}
\end{equation*}
Tensor products are associative and commutative, hence %
\begin{equation*}
\begin{split}
\mbb{C}\otimes_\mbb{R}(\mrm{G}\otimes \mrm{H})%
&=(\mbb{C}\otimes_\mbb{R}\mrm{G})\otimes \mrm{H}
=\mrm{G}\otimes(\mbb{C}\otimes_\mbb{R}\mrm{H})
.
\end{split}
\end{equation*}
\end{definition}

\begin{lemma}[\ref{Lem:12}]
\begin{equation}
\GL{ab}{R}\cong\GL{a}{R}\otimes\GL{b}{R}.\label{Lem:12}
\end{equation}
\end{lemma}
\begin{lproof}
The elements of $\gl{a}{R}$ take their values on the vector space defined by basis $\{e_{ij}\}$ over the real field. The product $\gl{a}{R}\otimes\gl{b}{R}$ admits a basis
\begin{equation*}
\left\{e_{mn}~\bigg|~\begin{tabular}{l}$e_{ij}\in\gl{a}{R},~e_{kl}\in\gl{b}{R},$\\$m=b(i-1)+k,~n=b(j-1)+l$\end{tabular}\right\}
\end{equation*}
which is identical to the $e_{ij}$ basis of $\gl{ab}{R}$, and this representation of $\gl{a}{R}\otimes\gl{b}{R}$ inherits by construction a bracket operation equivalent to that on $\gl{ab}{R}$. %
The vector spaces are isomorphic, the bracket operators are equivalent, so isomorphism of the Lie algebras follows,
\begin{equation*}
\gl{ab}{R}\cong\gl{a}{R}\otimes\gl{b}{R}.%
\end{equation*} 
This, in turn, implies local satisfaction of Lemma~\eref{Lem:12} on an open disc in the vicinity of the identity element. By Lie's third theorem---or more accurately, by Cartan's theorem, the infinitesimal of which is the inverse of Lie's third theorem---all locally isomorphic groups are either globally isomorphic or differ by a central extension, and therefore may be classified by their centres \cite{cartan1930,cartan1952,serre1992,gorbatsevich1986,tuynman2021}. The centre of $\GL{n}{R}$, denoted $c[\GL{n}{R}]$, is $D_n$, and thus
\begin{align}
c[\GL{a}{R}\otimes\GL{b}{R}]&=c[\GL{a}{R}]\otimes c[\GL{b}{R}]\nn\\
&\cong c[\GL{ab}{R}].
\tag{\ref{Lem:12}b}\label{eq:Lem12b}
\end{align}
Lemma~\eref{Lem:12} follows immediately. %
\qed
\end{lproof}

\begin{lemma}[\ref{Lem:1}]
\begin{equation}
\GL{ab}{C}\cong\GL{a}{R}\otimes\GL{b}{C}.\label{Lem:1}
\end{equation}
\end{lemma}
\begin{lproof}
Sequential application of Definition~\eref{Def:1}, Lemma~\eref{Lem:12}, and Definition~\eref{Def:1}. 
\qed
\end{lproof}

\begin{lemma}[\ref{Lem:7}]
\begin{equation}
\SL{n}{C}\cong\SU{n,\mbb{C}}.\label{Lem:7}
\end{equation}
\end{lemma}
\begin{lproof}
$\SU{n}$ is an $\mbb{R}$-form of $\SL{n}{C}$, corresponding to the fixed points of the conjugation operation
\begin{equation*}
\left.%
^\dagger:m\longrightarrow m'\qquad m'_{ij}=(m_{ji})^*
\quad\right|\quad m\in\SL{n}{C}.
\end{equation*}
That is,
\begin{equation*}
\begin{split}
\SU{n}&\cong\{g\in\SL{n}{C}~|~g^\dagger=g\}.
\end{split}
\end{equation*}
Complexification of any $\mbb{R}$-form recovers the parent group, and thus %
\begin{equation*}
\SU{n,\mbb{C}}:=\mbb{C}\otimes_\mbb{R}\SU{n}\cong\SL{n}{C}.
\end{equation*}
\qed
\end{lproof}

\begin{lemma}[\ref{Lem:9}]
\begin{equation}
\GL{2n}{C}\cong\left[\SU{n}\oplus\GL{1}{R}\right]\otimes\GL{2}{C}.\label{Lem:9}
\end{equation}
\end{lemma}
\begin{lproof}
Consecutive application of Lemmas~\eref{Lem:1}, \eref{Lem:4}, %
and~\eref{Lem:7}
yields
\begin{equation*}
\GL{2n}{C}\cong\left[\SU{n,\mbb{C}}\oplus\mbb{C}^\times\right]\otimes\GL{2}{R}.
\end{equation*}
Replacing $\mbb{C}^\times\cong\GL{1}{C}$ and applying Definition~\eref{Def:1} gives
\begin{equation*}
\GL{2n}{C}\cong\mbb{C}\otimes_\mbb{R}\left[\SU{n}\oplus\GL{1}{R}\right]\otimes\GL{2}{R}.
\end{equation*}
Further application of Definition~\eref{Def:1} %
completes the proof.\qed
\end{lproof}

\section{Gaugeable degrees of freedom in \protect{$\SU{3}$}\label{apdx:SU3gauge}}

To count the gaugeable degrees of freedom in the $\SU{3}$ local symmetry of the emergent model, recognise that the rescaled Gell-Mann matrices $\ulambda_1,\ldots,\ulambda_8$ \erefr{eq:Cbasis1}{eq:Cbasis2} comprise a basis for the Lie algebra to a faithful representation of $\SU{3}$, and that linear recombination of these matrices can yield 
\begin{align}
\begin{array}{ccc}
\tau^A_1=\frac{1}{\sqrt{2}}\bgrid 0&1&0\\1&0&0\\0&0&0 \egrid &
\tau^A_2=\frac{1}{\sqrt{2}}\bgrid 0&-\rmi&0\\\rmi&0&0\\0&0&0 \egrid \\
\tau^A_3=\frac{1}{\sqrt{2}}\bgrid 1&0&0\\0&-1&0\\0&0&0 \egrid \\
\end{array}\label{eq:taubasis1}
\end{align}
\begin{align}
\begin{array}{ccc}
\tau^B_1=\frac{1}{\sqrt{2}}\bgrid 0&0&0\\0&0&1\\0&1&0 \egrid &
\tau^B_2=\frac{1}{\sqrt{2}}\bgrid 0&0&0\\0&0&-\rmi\\0&\rmi&0 \egrid \\
\tau^B_3=\frac{1}{\sqrt{2}}\bgrid 0&0&0\\0&1&0\\0&0&-1 \egrid \\
\end{array}\label{eq:taubasis2}
\end{align}
\begin{align}
\begin{array}{ccc}
\tau^C_1=\frac{1}{\sqrt{2}}\bgrid 0&0&1\\0&0&0\\1&0&0 \egrid &
\tau^C_2=\frac{1}{\sqrt{2}}\bgrid 0&0&\rmi\\0&0&0\\-\rmi&0&0 \egrid \\
\tau^C_3=\frac{1}{\sqrt{2}}\bgrid -1&0&0\\0&0&0\\0&0&1 \egrid ,
\end{array}\label{eq:taubasis3}
\end{align}
where each triplet $\{\tau^X_1,\tau^X_2,\tau^X_3\}$ is the basis for a copy of $\SU{2}\subset\SU{3}$ denoted $\SU{2}_X$. However, the three diagonal elements $\tau^A_3$, $\tau^B_3$, and $\tau^C_3$ are not linearly independent since $\tau^C_3$ can be written as 
\begin{equation}
\tau^C_3=-\tau^A_3-\tau^B_3.\label{eq:tauC3}
\end{equation}
The remaining eight elements are linearly independent and form a basis for $\SU{3}$.

It is well-known that $\SU{2}$ has two gaugeable degrees of freedom, corresponding to a double cover of the Euler angles of the sphere \cite[e.g.{}][]{frankel2004}. The linearly independent subgroups $\SU{2}_A$ and $\SU{2}_B$ therefore contribute two gaugeable degrees of freedom apiece. If subgroup $\SU{2}_C$ was also linearly independent it would contribute a further two gaugeable degrees of freedom. However, it loses one degree of freedom through \Eref{eq:tauC3}. 
The (double cover of the) Euler angles parameterise the (double cover of the) unit 2-sphere in a three-dimensional parameter space. Fixing the coefficient of \Eref{eq:tauC3} restricts that parameter space to a two-dimensional subspace which includes the origin, and which intersects that (double cover of the) unit 2-sphere to yield a reduced gauge parameter space isomorphic to the (double cover of the) unit circle.
Thus
gauging $\SU{2}_A$ and $\SU{2}_B$ also fixes one of the gauge degrees of freedom of $\SU{2}_C$, leaving one further gauge degree of freedom unconstrained.

The number of independent gaugeable degrees of freedom on $\SU{3}$ is therefore five.

An alternative derivation, which does not depend on the identification of $\SU{2}$ subgroups, is to recognise that \Erefr{eq:Cbasis1}{eq:Cbasis2} comprise a basis for the Lie algebra to a faithful representation of $\SU{3}$, and that this representation acts on a complex-valued three-element vector having six degrees of freedom. Elements of $\SU{3}$ may map any such vector onto any other such vector, up to the constraint that they cannot modify the norm, so the dimension of the space spanned by the action of $\SU{3}$ on some initial vector is five. By construction this five-dimensional space is isomorphic to the parameter space of a local choice of gauge on $\SU{3}$.

\appendixend

\notchap{
\section*{Acknowledgements}
This research was supported in part by the Perimeter Institute for Theoretical Physics.
Research at the Perimeter Institute is supported by the Government of Canada through Industry Canada and by the Province of Ontario through the Ministry of Research and Innovation.
The author thanks the Ontario Ministry of Research and Innovation Early Researcher Awards (ER09-06-073) for financial support.
This project was supported in part through the Macquarie University Research Fellowship scheme.
This research was supported in part by the ARC Centre of Excellence in Engineered Quantum Systems (EQuS), Project No.~CE110001013.
}

\bibliographystyle{apsrev4-2}
\bibliography{Paper.bib}

\end{document}